\documentclass[final,3p,times,twocolumn]{elsarticle}
\biboptions{comma,sort&compress}
\usepackage{here}
\usepackage{graphicx}
\usepackage{epsfig}

\def\nin{\noindent}
\newcommand{\ima}{{\mbox{Im}\,}}
\newcommand{\ar}{\arrowvert}
\newcommand{\ra}{\rangle}

\newcommand{\ov}{\overline}

\newcommand{\be}{\begin{equation}}
\newcommand{\ee}{\end{equation}}
\newcommand{\bea}{\begin{eqnarray}}
\newcommand{\eea}{\end{eqnarray}}
\newcommand{\ba}{\begin{eqnarray}}
\newcommand{\ea}{\end{eqnarray}}

\journal{Nuc. Phys. (Proc. Suppl.)}
\begin{document}
\begin{frontmatter}



\title{Fock space expansion of $\sigma$ meson in leading-$N_c$}

 \author{Felipe J. Llanes-Estrada $^*$, Jose Ram\'on Pel\'aez and Jacobo Ruiz de Elvira}
  \address{Departamentos de F\'{\i}sica Te\'orica I y II, Universidad Complutense de Madrid, 28040 Madrid, Spain.}
\cortext[cor1]{Speaker}

\begin{abstract}
\noindent
We examine the  leading-$N_c$ behavior of the masses and transition matrix 
elements of some low-lying, few-particle configurations in QCD. 
A truncation of the Fock space produces an effective, symmetric 
Hamiltonian  that we diagonalize. The lowest eigenvalue  is identified as the 
$\sigma$ meson if the Hamiltonian is chosen to represent the scalar sector. 
As an application, the coefficients of the $N_c$ powers are then fit to 
two-loop Unitarized SU(2) Chiral Perturbation Theory results for the $\sigma$ mass 
and width as a function of the number of colors, 
and we show that those results can be accommodated using
the QCD $N_c$ dependence
previously derived for matrix elements,
without the need for unnatural parameters or fine tunings.
Finally, we show a very preliminary good quality fit, estimating
 the proportion of tetraquark/molecule-like (dominant), $q\bar{q}$-like (subdominant) 
and exotic-like (marginal) configurations in the $\sigma$.
\end{abstract}


\end{frontmatter}

\vspace{-.2cm}
\section{Motivation}
\vspace{-.2cm}

A low-mass $\sigma$ meson was introduced in 1955 \cite{Teller} 
as an auxiliary device that has turned out to be very useful to  explain the intermediate-distance attractive part in the nuclear potential.
For long time debated, the mass and width of this meson (decaying almost always to $\pi\pi$, the only open strong decay channel) have been recently pinned down with very good precision employing different methods. The results of these analysis are in agreement with each other and some are shown in table \ref{table:masswidth}.

{\scriptsize
\begin{table}[hbt]
\setlength{\tabcolsep}{1.4pc}
 \caption{\scriptsize Precise determinations of the $\sigma$ meson mass and half-width (in $MeV$).}
    {\small
\begin{tabular}{ccc}
&\\
\hline
$M$ & $\Gamma/2$ & Refs.    \\
\hline
452(12)  & 260(15) & \cite{Mennessier} \\
458(15)  & 262(15) & \cite{Ruben} \\
441(12)  & 272(12) & \cite{Caprini} \\
\hline
\end{tabular}
}
\label{table:masswidth}
\end{table}
}
\nin

It behooves one to understand the composition of this meson in terms of the fundamental QCD degrees of freedom, quarks and gluons. Of current interest is the decomposition of states in terms of a Fock space expansion \cite{Schechter} and this we address in the present brief report.
This expansion reads
\be
\ar \sigma \ra =\sum\int\left( \alpha_{q\bar{q}} \ar q\bar{q} \ra +
\alpha_{gg} \ar g\bar{g} \ra + \alpha_{qq\bar{q}\bar{q}} \ar qq\bar{q}\bar{q} \ra \dots \right) 
\ee
where the sum/integral signs remind us of spin, momentum and other degrees of freedom 
that, for simplicity, we will further omit in our notation;
that is, we consider the summed amplitude over each Fock subspace  
\be
\ar \sigma \ra = \alpha_{q\bar{q}} \ar q\bar{q} \ra +
\alpha_{gg} \ar g\bar{g} \ra + \alpha_{qq\bar{q}\bar{q}} \ar qq\bar{q}\bar{q} \ra \dots 
\label{Fock2} 
\ee
This expansion in terms of quarks and transverse gluons is well defined in Coulomb gauge QCD~\cite{Adam}, that can be formulated without ghosts nor longitudinal gluons. At least for heavy mesons decaying to open-flavor channels, the intrinsic $q\overline{q}$ component can be identified in a model-independent way~\cite{TorresRincon}. The setback of this full quantum-mechanical answer is that it is frame-dependent, presumably defined in the rest frame of the hadron~\cite{Rocha}. This makes it less attractive for light hadrons where speeds can be large.

The $1/N_c$ expansion around $N_c=3$ offers more limited information: it can only separate classes of equivalence of states whose mass and decays behave in the same way under $N_c$, but the information obtained is useful also for light quarks.
Thus our states in Eq.(\ref{Fock2})
should be understood as $q\bar{q}$-like, $g\bar{g}$-like, etc... although 
for simplicity we are calling them $q\bar{q}$, $g\bar{g}$, etc...

\vspace{-.2cm}
\section{Matrix elements in leading-$N_c$}
\vspace{-.2cm}

We now consider what configurations may play an important role for the $\sigma$ meson (and other light scalar  mesons). First, the $\sigma$ is a very broad resonance in $\pi\pi$ scattering, deep in the complex plane, so one can describe part of its nature as a pion-pion correlation (distortion of the density of states, or for brevity, `molecule'), or equivalently as the leading-$N_c$ color analysis is concerned, a tetraquark. 
Thus, although we denote it by
$q\bar{q}q\bar{q}$, we actually mean ``tetraquark/molecule''.

Next, at some level one expects to find (and indeed finds as will be shown in figure~\ref{SigmaMass} below) a $q\ov{q}$ component, that would correspond to the $1\ $GeV quark model's $ ^3P_0$ configuration. 

Finally, one might expect more exotic configurations such as glueballs or baryonium-type multiquark correlations to also play a role.

We give the leading $N_c$ behavior, $N_c^\beta$,
of the mass and $\pi\pi$ width of 
these various configuration in table~\ref{table:ncbehavior}.

\begin{figure}[hbt] 
\centerline{\includegraphics[width=7.5cm]{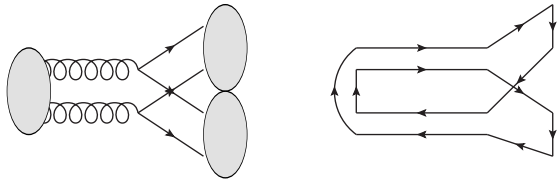}}
\caption{\scriptsize Left: the impulse diagram for the transition of a glueball to two pions already yields the leading-$N_c$ behavior of the entire amplitude as shown by t'Hooft. Right: color flow of the same diagram. The line crossing reveals the $1/N_c$ suppression.}
\label{Ncbehavior} 
\end{figure}

{\scriptsize
\begin{table}[hbt]
\setlength{\tabcolsep}{1.4pc}
 \caption{\scriptsize Leading-$N_c$ scaling of mass and width of various QCD Fock-states.}
    {\small
\begin{tabular}{ccc}
&\\
\hline
State      & $M$    & $\Gamma_{\pi\pi}$    
\\ \hline
$\pi\pi$, $q\bar{q}q\bar{q}$ & $O(1)$ & O(1) 
\\
$q\bar{q}$ & $O(1)$ & $O(1/N_c)$ 
\\ 
$gg$       & $O(1)$ & $O(1/N_c^2)$
\\
$(N_c-1)(q\bar{q})$ & $O(N_c)$ & $O(e^{-N_c})$
\\
\hline
\end{tabular}
}
\label{table:ncbehavior}
\end{table}
}
\nin
We illustrate the color computation of one of these matrix elements (the glueball to two pion transition $G\to \pi\pi$) in figure \ref{Ncbehavior}.
\nin

A way to establish the counting (left diagram in the figure) is to observe that the color-singlet two-gluon wavefunction, properly normalized, is
$\frac{\delta_{ab}}{\sqrt{N_c^2-1}}$. Each of the two vertices carry $\frac{gT^a_{ij}}{\sqrt{N_c}}$ (this scaling of the color charge guarantees that higher-order diagrams scale in the same way under $N_c$). Finally, the pion wavefunctions in the final state combine a quark and antiquark to form a color singlet $\frac{\delta_{ij}}{\sqrt{N_c}}$. 

The net result for the matrix element is $tr(T^a T^a)/(N_c^2\sqrt{N_c^2-1})$,
suppressed as $1/N_c$.  As the width is proportional to the matrix element $G\to \pi\pi$ squared, the corresponding entry in table \ref{table:ncbehavior}  follows.

\vspace{-.2cm}
\section{Effective Hamiltonian}
\vspace{-.2cm}

We take one state from each class $q\ov{q}$, $gg$, $q\ov{q}q\ov{q}$ \footnote{
For this analysis we discard the baryonium-like configuration, that will be reexamined in an upcoming work.}, to build a discrete $3\times 3$ effective Hamiltonian $H$. We will not attempt to calculate this Hamiltonian from theory, but we factor the leading-$N_c$ behavior of its matrix elements, known from a $N_c$ analysis, and leave the pre-coefficients as free parameters.
Diagonalization of $H$ yields three eigenvalues. We identify the lowest one as the $\sigma$.
 
To describe the width, our three model states have to be coupled to the pion-pion continuum. For this we employ a Feshbach decomposition~\cite{Feshbach} in terms of a $P$ subspace (our three discrete states) and a $Q$ subspace (two free pions in an arbitrary relative momentum state). The full Hamiltonian in the total space
\be
H= \left(
\begin{tabular}{cc}
$H_{PP}$ & $H_{PQ}$\\
$H_{QP}$ & $H_{QQ}$
\end{tabular}
\right)  
\ee
is then restricted to the discrete $P$ subspace via the resolvent in 
$Q$-space with appropriate boundary conditions 
\be \label{Heff}
H^{eff}_{PP} = H_{PP} + H_{PQ} \frac{1}{E-H_{QQ}+i\epsilon} H_{QP}\ .
\ee

We actually do not need to calculate the integral over pion configurations in the rightmost term; all we need is to extract its leading-$N_c$ behavior. 
The effective Hamiltonian is finally a symmetric (because of $CP$ invariance), non-Hermitian (because of the Fock-space restriction) $3\times 3$ complex matrix, that has therefore 12 free parameters in leading $N_c$ (the exponents being known),
\be
H_{ij}= h_{ij} \times N_c^{\beta_{ij}} \ .
\ee
The diagonal $\beta$ are given in Table 2, but for brevity,
all others -- already calculated with the procedure
explained above--  will be given somewhere else.
We express the lowest eigenvalue (now complex) in terms of these unspecified parameters $h_{ij}$.

\vspace{-.2cm}
\section{Inverse Amplitude Method}
\vspace{-.2cm}

\begin{figure}[hbt]
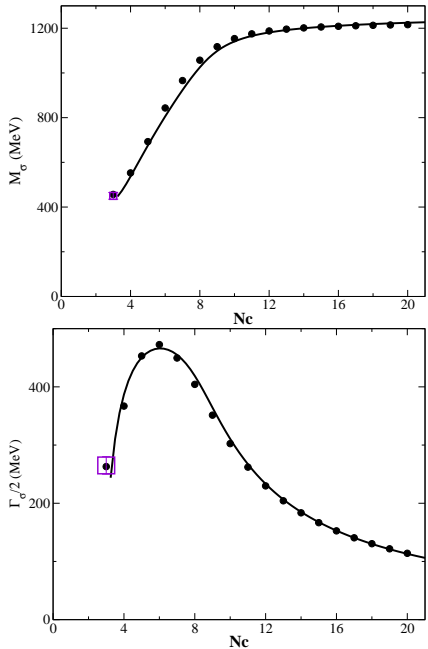
 
\centerline{\includegraphics[width=5.5cm]{msigma.eps}}
\centerline{\includegraphics[width=5.5cm]{wsigma.eps}}
\caption{\scriptsize Mass (top) and width (bottom) of the $\sigma$ meson in Unitarized Chiral Perturbation Theory (dots) as a function of the number of colors. The solid line is the fit to these data points in the three-state
model that includes a molecule-like state, and two intrinsic $q\overline{q}$ and $gg$ states. Squares (purple colored online) at $N_c=3$ represent the precise dispersive determinations from table~\ref{table:masswidth}.}
\label{SigmaMass} 
\end{figure} 
\nin

We match our $N_c$ results to a dispersive analysis, the well-established Inverse Amplitude Method (IAM) \cite{Truong,Dobado}. The method
proceeds by writing a dispersion relation for  $G\equiv \frac{(t^{(2)})^2}{t}$ where $t$ is the scalar, isoscalar pion-pion scattering amplitude and
$t=t^{(2)}+t^{(4)}+t^{(6)}\dots$ is its ChPT expansion. This reads
\begin{eqnarray}\nonumber  
G(s)&=&G(0)+G'(0)s+\frac{1}{2}G''(0)s^2
\\ \nonumber  
&+&\!\!\!\!\frac{s^3}{\pi}\int_{RC}ds'\frac{\ima G(s')}{s'^3(s'-s)}+LC(G)+PC(G)
\end{eqnarray}
in terms of a left Mandelstam cut (LC), pole contributions (PC), due to the Adler
zeroes of $t$ and analyzed elsewhere~\cite{GomezNicola:2007qj}, polynomial subtractions to assist convergence, and a dispersive integral over the right, unitarity cut.
The two first terms in the first line are approximated in Chiral Perturbation
Theory. Their contribution to the amplitude in the physical region with $E\ge 2m_\pi$ is very small since they lay well to the left of this threshold in the complex $s$-plane. The polynomic subtraction is represented exactly in the chiral expansion. 

This leaves the dispersive integral on the second line. 
The nice feature about it is that, in the elastic region for $\pi\pi$
scattering, with $2m_K\ge E\ge 2m_\pi$, the imaginary part is exactly known,
$\ima G = -\ima t_4$, and this formula provides a good
approximation further up to $E\simeq 1.2\ $ GeV.

The dispersion relation can then be turned into simple algebraic expressions to the order desired, and while the approximation to order $p^4$ has been much exploited, we here employ the order $p^6$ in the expansion, that yields  (not writing down the pole contributions for simplicity)
\be \label{amplitude}
t\simeq \frac{t_2^2}{(t_2-t_4+t_4^2/t_2-t_6)}\ .
\ee

The poles (elastic resonances) in pion-pion scattering are thus simply obtained 
as zeroes of the denominator, and the $\sigma$ mass and width extracted thereof.
The dependence $N_c$ dependence of the resonance parameters 
then obtained  \cite{Pelaez:2003dy}\cite{Guillermo}
by changing the chiral low energy parameters following
 their model independent ChPT/QCD description  \cite{Gasser:1984gg}
(that follows from counting flavor traces).
For this $\sigma$ meson application we need only
$f_\pi \to f_\pi\sqrt{\frac{N_c}{3}}$, $l_i \to l_i \frac{N_c}{3}$ for
$i=1\dots 4$ and $r_i \to r_i \left(\frac{N_c}{3}\right)^2$ for
$i=1\dots 6$.  The values of the low energy constants at $N_c=3$ are given in table \ref{Tab:SU2lecs}.
\begin{table}
  \begin{tabular}{cc|cc}\hline
    $l_1^r$(x $10^3$) & -5.4 &   $r_1$(x $10^4$)  & -0.6\\
    $l_2^r$(x $10^3$) & 1.8  &   $r_2$(x $10^4$)  & 1.5 \\
    $l_3^r$(x $10^3$) & 1.5  &   $r_3$(x $10^4$)  & -1.4\\
    $l_4^r$(x $10^3$) & 9.0  &   $r_4$(x $10^4$)  & 1.4 \\ 
                      &      &   $r_5$(x $10^4$)  & 2.4 \\
                      &      &   $r_6$(x $10^4$)  & -0.6\\ \hline 
  \end{tabular}\caption{Two-loop IAM LECs employed, corresponding to the
    fit $\rho$ as $\bar{q}{q}$ in \cite{Guillermo}. We rely on $SU(2)$ chiral perturbation theory, while it is known that the subthreshold coupling of the $\sigma$ to the closed $K\overline{K}$ channel is large~\cite{Mennessier}. A pure color analysis is blind to these flavor details, and for the time being the $K\ov{K}$ component must be understood as included in the (dominant) meson-meson component analyzed.}\label{Tab:SU2lecs}
\end{table}

Figure \ref{SigmaMass} does not exceed $N_c\simeq 20$. The IAM is reliable near $N_c=3$ where the resummation of $s$-channel rescattering effected by the IAM is the dominant physics, and the unitarity cut dominates the dispersion relation. Also a flavor singlet Goldstone boson is not necessary in the effective Lagrangian for modest $N_c$, as the axial anomaly is $N_c$ suppressed. Our results should be understood as a {\it{leading}}-$N_c$ expansion around $N_c=3$,
{\it never as $N_c\rightarrow\infty$}, see~\cite{Arriola}.

\vspace{-.2cm}
\section{Results}
\vspace{-.2cm}

The IAM results~\cite{Guillermo} for varying-$N_c$ are displayed in figure \ref{SigmaMass}. The mass $M_\sigma$ initially increases towards $1\ $GeV with growing $N_c$,
but then saturates to a more or less constant value after $N_c\simeq 8$. The
half-width $\Gamma_\sigma/2$ also starts increasing with $N_c$, but around
$N_c\simeq 6$, it decreases as it is expected for a $q\ov{q}$ meson. This
behaviour has been interpreted~\cite{Guillermo} as a signal of the onset of the intrinsic,
$q\ov{q}$ subdominant component of the $\sigma$ meson and seems to be needed to ensure fulfillment of local duality \cite{Jacobo}. 
The behavior of the width at lower $N_c$ is characteristic of a molecule or tetraquark component.

It is plain from a comparison with table~\ref{Ncbehavior} that none of the intrinsic QCD states by itself can reproduce this behavior. One therefore needs to consider the mixing between different configurations.

Employing our $3\times 3$ effective Hamiltonian, 
and varying $N_c$ in the known manner, we fit the free pre-coefficients $h_{ij}$. 
We restrict the fit to the parameter subspace yielding one light scalar only 
(in agreement with Unitarized Chiral Perturbation Theory), the other two being 
above $1.2\ $GeV, and we make no further statement about them since at that 
point they escape the reach of our method.
We further examine naturality in the $N_c$ pre-coefficients, 
in the sense that a coefficient of the known $N_c$ powers is said to be natural if 
$h_{ij}\in (1/N_c,N_c)$.

This equation guarantees that the naive $N_c$ counting works for the 
effective Hamiltonian. It could accidentally happen that one of the pre-coefficients of a subleading term was very large and the lowest orders in the $1/N_c$ expansion were therefore not a reliable approximation. 
Our naturality assumption means that we discard this case and believe the $N_c$ counting {\it{as is}}.  

\begin{figure}[hbt] 
\centerline{\includegraphics[width=6.5cm]{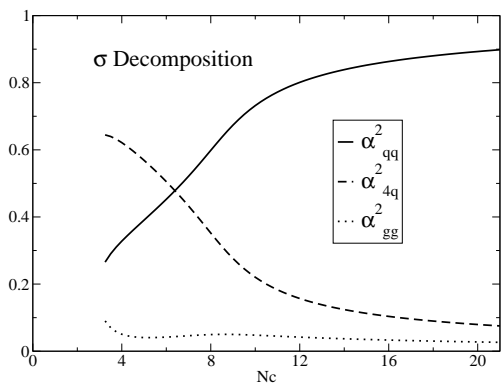}}
\caption{\scriptsize Fock space decomposition of the $\sigma$ meson in a discrete model 3-state subspace, imposing naturality for the effective $3\times 3$ Hamiltonian, whose pre-coefficients are obtained fitting leading-$N_c$ to Unitarized Chiral Perturbation Theory.  The dominant component behaves under $N_c$ as a $q\ov{q}q\ov{q}$ through $N_c\simeq 6$, then the subdominant $q\ov{q}$-like takes over for larger $N_c$. The glueball-like component stays always at or below the $10\%$ level.}
\label{SigmaCoeffs} 
\end{figure}

Our very preliminary best fit is shown in figure~\ref{SigmaCoeffs}, 
where we plot the probabilities $\ar \alpha\ar^2$ of finding one of our three model 
states in the lowest eigenvector.

The graph encodes all information that can be extracted from the $N_c$ counting and naturality alone without increasing the model space. Given the large number of parameters in the minimization, we would be cautious about extending $P$ beyond, say, four states.

It is rewarding that a good fit can be found (the solid line in figure~\ref{SigmaMass}) with such a simple model.

One could also inquire whether alternative fits that make the exotic (glueball) component dominant in the $\sigma$ expansion could be found as some authors indicate that this might be the case~\cite{Mennessier}. This is ongoing work that will be reported in a follow-up publication.

A similar analysis could also be carried out for baryons and in fact, another problematic state in the low hadron spectrum, the $\Lambda(1405)$, is being examined~\cite{Hosaka}.

\vspace{-.2cm}
\section*{Acknowledgements}
\vspace{-.2cm}
\nin
We thank S. Narison for the invitation to the inspiring 2010 edition of this 
conference, and interesting discussions with R.~Jaffe, J.~A.~Oller. Work supported by grants FPA2008-00592, FIS2008-01323, FIS2006-03438 (MICINN), U.Complutense/Banco Santander grant PR34/07-15875-BSCH and UCM-BSCH GR58/08 910309 and the EU-Research Infrastructure Integrating Activity ``Study of Strongly Interacting Matter'' (HadronPhysics2, Grant 227431) under the EU 7th Framework Programme.

\vspace{-.2cm}

\end{document}